\documentclass[sigconf, 9pt]{acmart}
\usepackage{blindtext}
\usepackage[utf8]{inputenc}
\usepackage{epsfig,xspace,url,xcolor}
\usepackage{amsmath}
\usepackage{mathtools}
\usepackage{amsthm}
\usepackage{tikz}
\usetikzlibrary{positioning,calc}
\usepackage{caption}

\usepackage{graphicx}
\usepackage[inline]{enumitem}
\usepackage{tikz}
\usepackage{thmtools}
\usepackage{thm-restate}
\usepackage{scalerel}
\usepackage[ruled, lined, linesnumbered, commentsnumbered, noend]{algorithm2e}
\usepackage[noend]{algpseudocode}
\usepackage{graphicx}
\usepackage{blindtext}
\usepackage{url}
\usepackage{subcaption}
\usepackage{mathtools}
\usepackage{xcolor}
\usepackage{color, colortbl}
\usepackage{tablefootnote}
\usetikzlibrary{positioning, shapes.geometric, shadows}
\usepackage[subtle]{savetrees}
\usepackage{graphbox}
\usepackage{url}

\usepackage[normalem]{ulem}
\usepackage{wrapfig}
\usepackage{wrapstuff}
\usepackage{breakurl}

\let\ACMmaketitle=\maketitle
\renewcommand{\maketitle}{\begingroup\let\footnote=\thanks \ACMmaketitle\endgroup}
\let\height\relax

\acmYear{2026}\copyrightyear{2026}
\setcopyright{cc}
\setcctype[4.0]{by-nc-nd}
\acmConference[HotOptics '26]{ACM SIGCOMM Workshop on Hot Topics in Optical Technologies and Applications in Networking}{August 17--21, 2026}{Denver, CO, USA}
\acmBooktitle{ACM SIGCOMM Workshop on Hot Topics in Optical Technologies and Applications in Networking (HotOptics '26), August 17--21, 2026, Denver, CO, USA}
\acmDOI{10.1145/3789240.3822610}
\acmISBN{979-8-4007-2467-1/26/08}



\usepackage[framemethod=TikZ]{mdframed}

\usepackage{enumitem}

\definecolor{darkgreen}{rgb}{0.01, 0.75, 0.24}
\definecolor{darkcyan}{rgb}{0.0, 0.55, 0.55}
\definecolor{mycolor}{rgb}{0.01, 0.75, 0.24}

\definecolor{mygray}{gray}{0.9}
\definecolor{myred}{rgb}{0.75, 0, 0}

\definecolor{myblue}{HTML}{3ba9d1}
\definecolor{myyellow}{HTML}{d1c23b}
\definecolor{mydarkred}{HTML}{c27676}
\definecolor{mygreen}{HTML}{7dc276}

\definecolor{alg1}{rgb}{0.855, 1, 0.824}
\definecolor{alg2}{rgb}{1, 0.996, 0.824}
\definecolor{alg3}{rgb}{0.824, 0.914, 1}

\mathchardef\hyphen="2D

\newcommand{\myitem}[1]{\vspace{1mm}\noindent\textbf{#1}}

\newcommand{\remove}[1]{}

\usepackage{xcolor}
\definecolor{light-gray}{gray}{0.9}
\usepackage{tcolorbox}

\newcommand{\ie}{i.e., \@}

\newcommand{\cref}[1]{Chapter~\ref{#1}}

\makeatletter
\newcommand{\oset}[2]{%
  \mathop{#2}\limits^{
  \vbox to-.5\ex@{\kern-2\ex@
   \hbox{$\scriptstyle#1$}\vss}}}
\makeatother

\DeclareSymbolFont{extraup}{U}{zavm}{m}{n}
\DeclareMathSymbol{\varheart}{\mathalpha}{extraup}{86}
\DeclareMathSymbol{\vardiamond}{\mathalpha}{extraup}{87}

\newcommand{\takeaway}[2]{{\textcolor{black}{$\blacksquare$ \textbf{\textit{Takeaway:}}} \textit{#1}}}
\title{Birkhoff Decompositions and Photonic Interconnects\\Wait! Don't Forget the Compute!
}

\author{Eliezer Amponsah}
\affiliation{
  \institution{Purdue University}
  \country{USA}
}

\author{Vamsi Addanki}
\affiliation{
  \institution{Purdue University}
  \country{USA}
}

\renewcommand{\shortauthors}{Amponsah et al.}

\begin{document}

\begin{abstract}
The growing demand for efficient communication in distributed training and inference has sparked significant interest in reconfigurable photonic interconnects across both academia and industry. Mixture-of-Experts (MoE) models, with their highly skewed communication patterns, present a natural opportunity for such circuit-switched fabrics. However, existing approaches largely optimize communication in isolation, overlooking the interaction between communication and the expert computation that follows.

In this paper, we revisit circuit scheduling for all-to-all communication in MoE execution. We show that the dispatch--compute--combine structure fundamentally challenges classical scheduling techniques such as Birkhoff--von Neumann (BvN) decomposition. First, MoE communication matrices are rarely doubly stochastic, introducing significant scheduling bubbles in BvN-based schedules. Second, while decomposition enables communication--compute overlap, the excessive number of matchings produced by BvN fragments execution into small batches, leading to severe compute inefficiencies due to fixed execution overheads. Motivated by these observations, we explore a simple greedy max-weight decomposition strategy that bounds the number of matchings while preserving large batch sizes per matching. Despite its simplicity, the approach significantly improves overlap efficiency, reduces compute overheads, and approaches the performance of an ideal congestion-free all-to-all.
\end{abstract}

\maketitle
\pagestyle{empty}
\thispagestyle{empty}

\section{Introduction}
\label{sec:intro}

Optical circuit switching has emerged as a promising technology for next-generation datacenter networks~\cite{10.1145/3387514.3406221,10.1145/3098822.3098838,10.1145/3651890.3672273,10.1145/3696348.3696856,10.1145/3651890.3672248,10.1145/3579312,torrijos2026industry,295551,10.1145/2486001.2486007,10.1145/1851182.1851223,10.1145/3708980,NANCEHALL2021100621}. Compared to traditional electronic packet switching, optical circuits offer substantially higher bandwidth and lower latency, making them particularly attractive for communication-intensive workloads such as large-scale distributed training and inference~\cite{10.1145/3696348.3696856,295551,10254691,10.1145/3772356.3772395}. However, these benefits come with an important constraint: a circuit-switched network can only realize a limited communication pattern at a given time and must repeatedly reconfigure itself to support changing traffic demands. Consequently, a significant body of work has focused on topology reconfiguration and circuit scheduling algorithms for efficiently mapping dynamic communication workloads onto optical fabrics~\cite{10.1145/3748273.3749203,10.1145/3748273.3749210,rahman2026harvestadaptivephotonicswitching,10.1145/3696348.3696856,10.1145/3712285.3759842,10.1145/3718958.3750465,10.1145/2716281.2836126,10.1145/2896377.2901479,10.1145/3409964.3461786, Huang2016Sunflow:Coflows}.

At the same time, modern AI workloads are becoming increasingly communication dominated~\cite{10.1145/3718958.3750465}. In particular, Mixture-of-Experts (MoE) architectures introduce highly irregular all-to-all communication due to dynamic token routing across distributed experts~\cite{Go2025MoETuner:Routing,Zhou2022Mixture-of-ExpertsRouting}. Unlike traditional collective communication patterns, MoE traffic matrices are often sparse and skewed, varying significantly across iterations and inputs. This makes MoE workloads a natural fit for reconfigurable optical fabrics capable of adapting the topology to the communication demand.

A common approach in circuit scheduling is to decompose a traffic matrix into a sequence of matchings that can be executed over time, often using techniques such as Birkhoff--von Neumann (BvN) decomposition~\cite{BIRKHOFF1946TresLineal,Wu2026DynamicCommunication}. At first glance, such decompositions appear attractive for distributed MoE execution: each matching directly translates into a communication schedule between GPUs, naturally limiting contention while enabling the network to adapt to the routing pattern of the current batch. Furthermore, decomposition implicitly partitions the all-to-all into smaller communication phases, creating opportunities to overlap communication with expert computation.

However, our study reveals that decomposition quality for communication is fundamentally different from decomposition quality for end-to-end MoE execution. In particular, the interaction between communication scheduling and downstream expert computation plays a central role in determining overall performance. While fine-grained decompositions improve network utilization and reduce contention, they also fragment execution into increasingly smaller token batches. In practice, expert execution exhibits substantial fixed overheads at small batch sizes, causing compute efficiency to collapse despite improved communication structure. As a result, decomposition strategies that appear favorable from a purely networking perspective can significantly degrade end-to-end makespan.

This tension is particularly pronounced for the widely studied Birkhoff--von Neumann (BvN) decomposition. Because MoE communication matrices are rarely doubly stochastic, BvN often produces decompositions with a large number of matchings and substantial idle capacity. The resulting execution suffers not only from additional reconfiguration overheads, but also from poor compute efficiency due to excessively fragmented expert batches. Our results show that, in many settings, minimizing communication congestion alone is insufficient and can even be counterproductive.

Motivated by these observations, we instead view distributed MoE execution as a scheduling problem that jointly considers communication decomposition and compute efficiency. Rather than pursuing highly fragmented decompositions, our goal is to preserve large execution batches while still exposing opportunities for communication--compute overlap. We show that a simple max-weight decomposition strategy is remarkably effective in this setting, substantially reducing execution makespan. In contrast, highly fragmented decompositions such as BvN suffer significant performance degradation due to poor compute efficiency at small batch sizes. By preserving larger execution batches, the max-weight approach exposes additional overlap opportunities between communication and expert computation, allowing it in many settings to approach and even outperform an idealized congestion-free all-to-all execution.

Motivated by these observations, our goal is to preserve large execution batches while maintaining communication--compute overlap. We show that a max-weight decomposition strategy is remarkably effective in reducing execution makespan. In contrast, BvN suffers significant performance due to poor compute efficiency at small batch sizes. The max-weight approach exposes additional overlap opportunities between communication and expert compute, allowing it to approach and even outperform an idealized congestion-free all-to-all execution.

We validate these observations using trace-driven simulations across various MoE workloads and decomposition strategies. Our results demonstrate that decomposition granularity, rather than communication efficiency alone, is a first-order concern in distributed MoE execution over reconfigurable optical fabrics.

\begin{figure}
\centering
\includegraphics[width=0.8\linewidth]{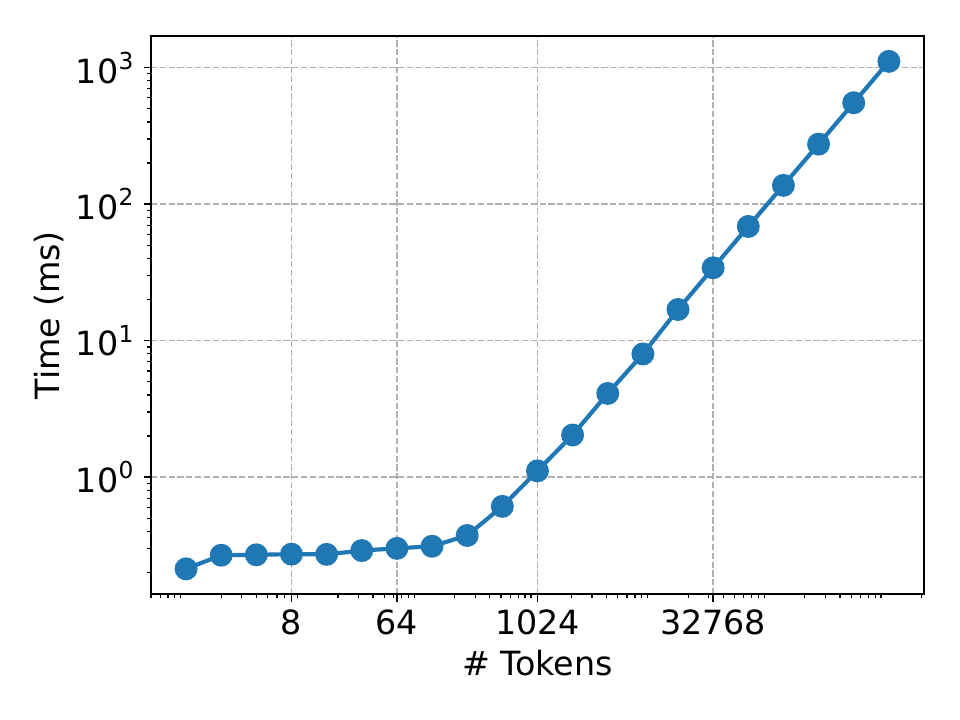}
\caption{MoE expert compute time across token batch sizes typically exhibits a ``knee'' behavior. While execution remains approximately linear beyond $256$ tokens, smaller batches incur substantial fixed overheads, causing arbitrary-size BvN decompositions to significantly underperform.}
\label{fig:compute-times}
\Description{Visual plot describing the "knee" behavior of compute times as the number of tokens increase. It shows a relatively constant compute time up to $\approx256$ token counts. After $\approx256$ tokens, the computation time observes a linear growth indicating that compute is dominated by fixed ovevrheads for small batches.}
\end{figure}

\begin{figure*}
\centering
\begin{subfigure}{0.138\linewidth}
\centering
\includegraphics[width=1\linewidth]{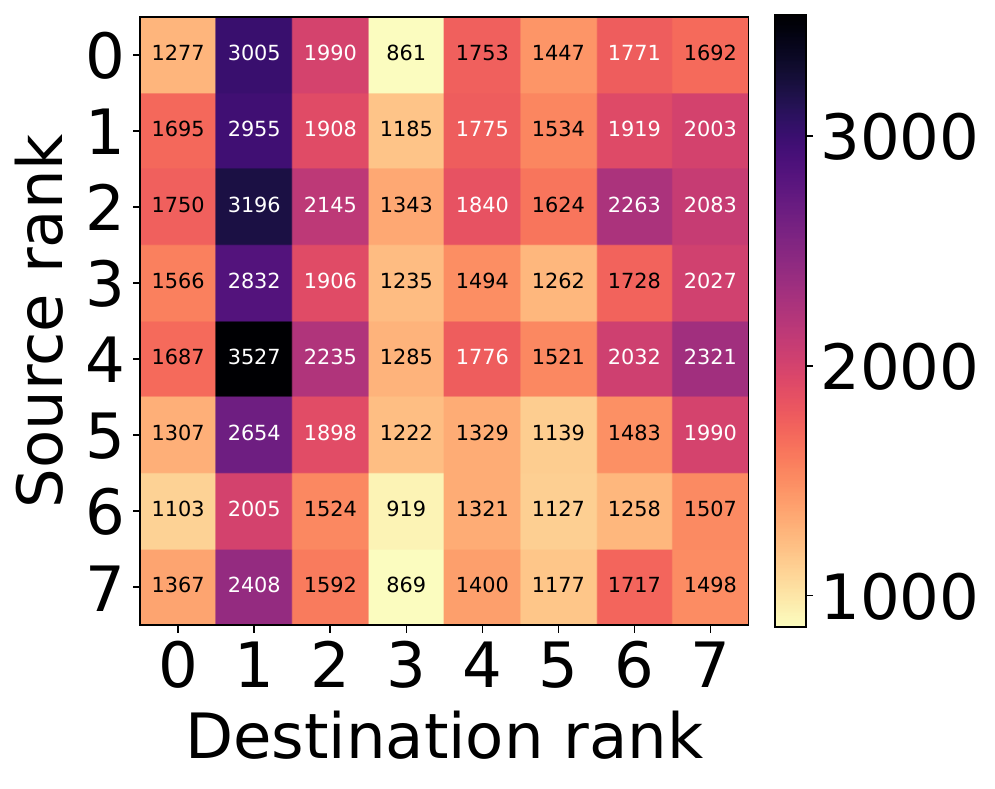}
\caption{Original A2A}
\end{subfigure}\hfill
\begin{subfigure}{0.138\linewidth}
\centering
\includegraphics[width=1\linewidth]{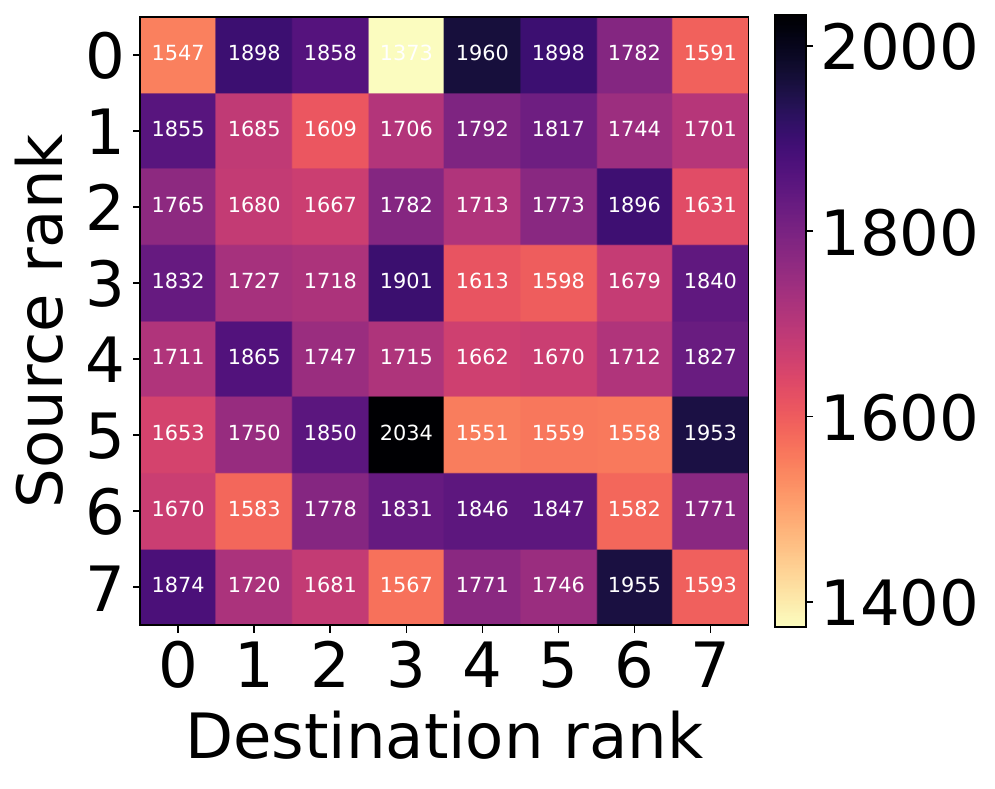}
\caption{Sinkhorn}
\end{subfigure}\hfill
\begin{subfigure}{0.138\linewidth}
\centering
\includegraphics[width=1\linewidth]{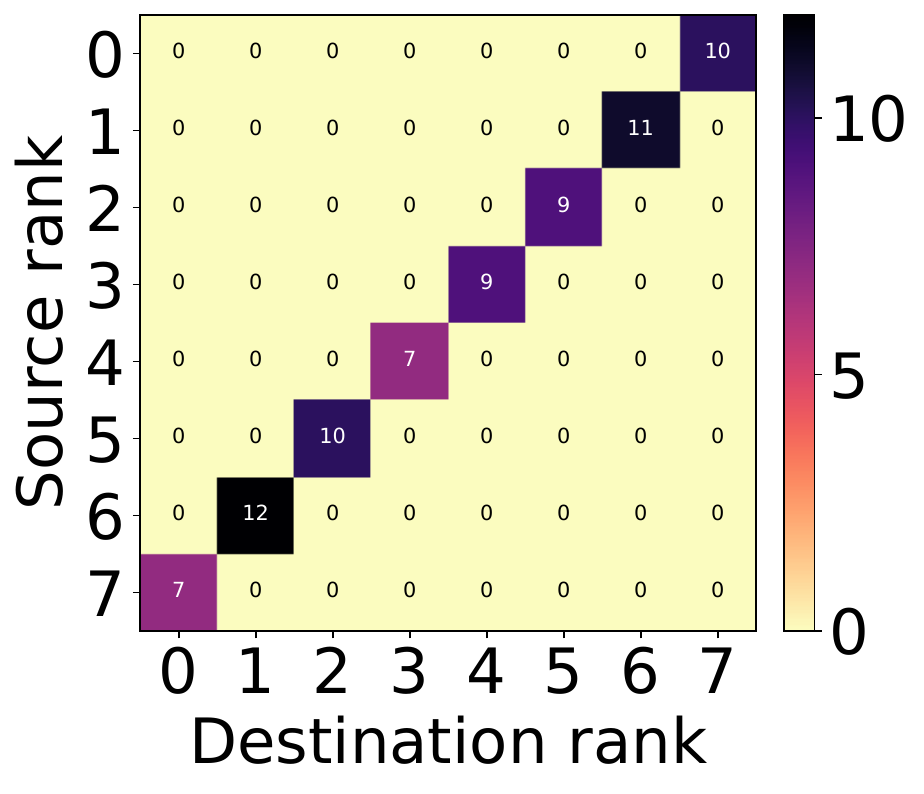}
\caption{BvN-$1/50$}
\end{subfigure}\hfill
\begin{subfigure}{0.138\linewidth}
\centering
\includegraphics[width=1\linewidth]{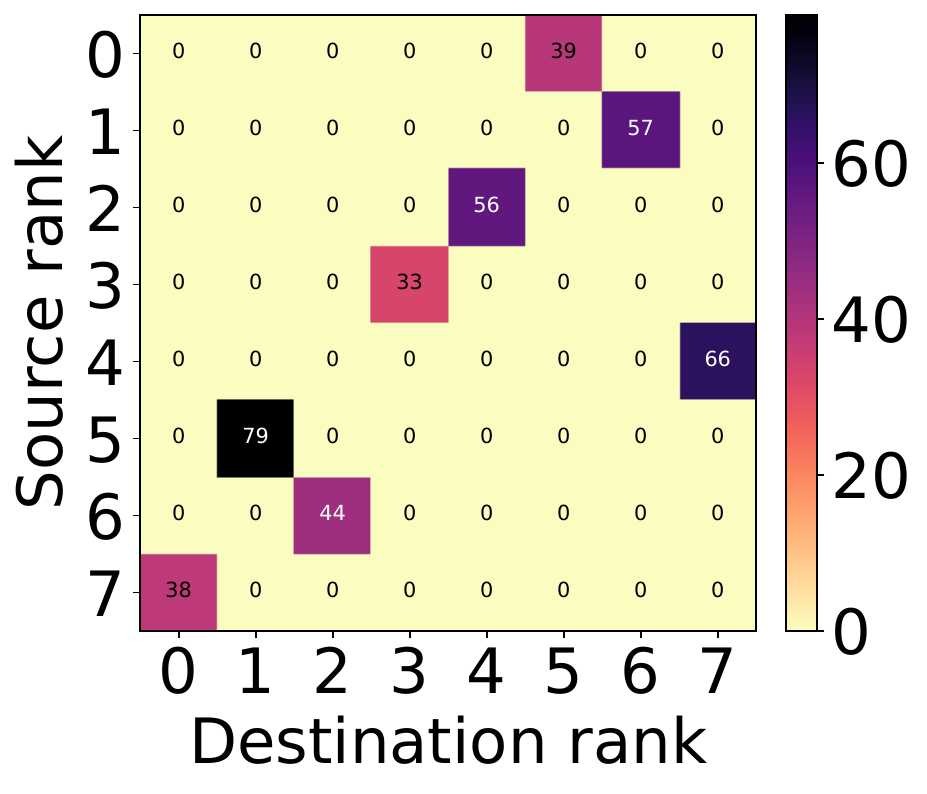}
\caption{BvN-$2/50$}
\end{subfigure}\hfill
\begin{subfigure}{0.138\linewidth}
\centering
\includegraphics[width=1\linewidth]{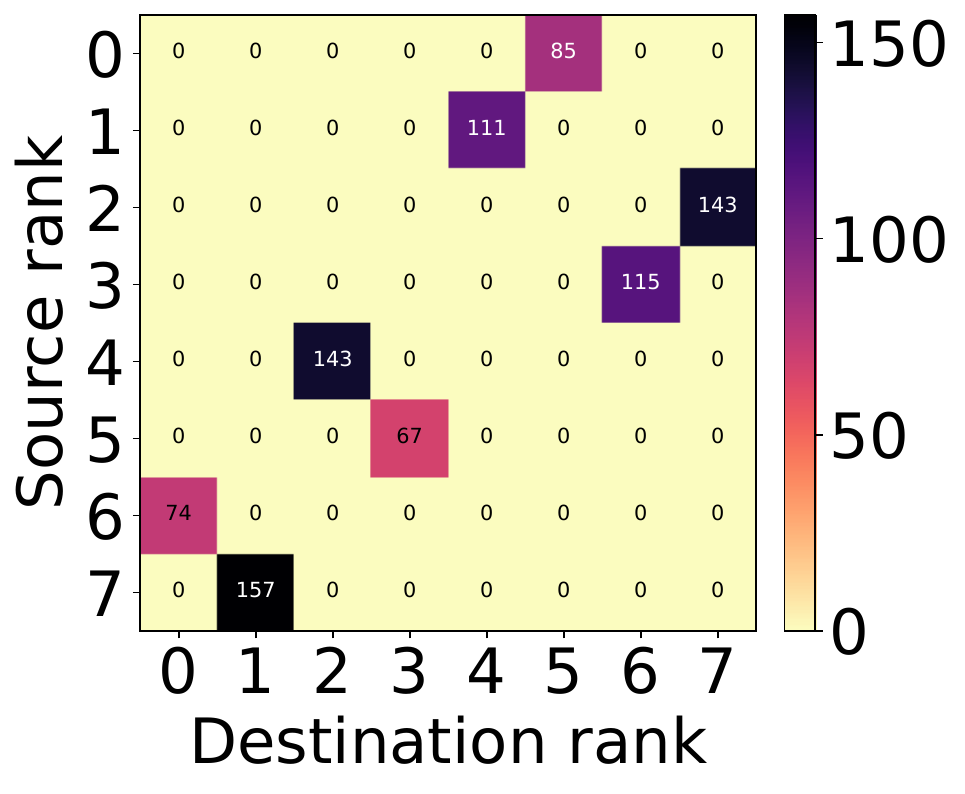}
\caption{BvN-$3/50$}
\end{subfigure}\hfill
\begin{subfigure}{0.138\linewidth}
\centering
\includegraphics[width=1\linewidth]{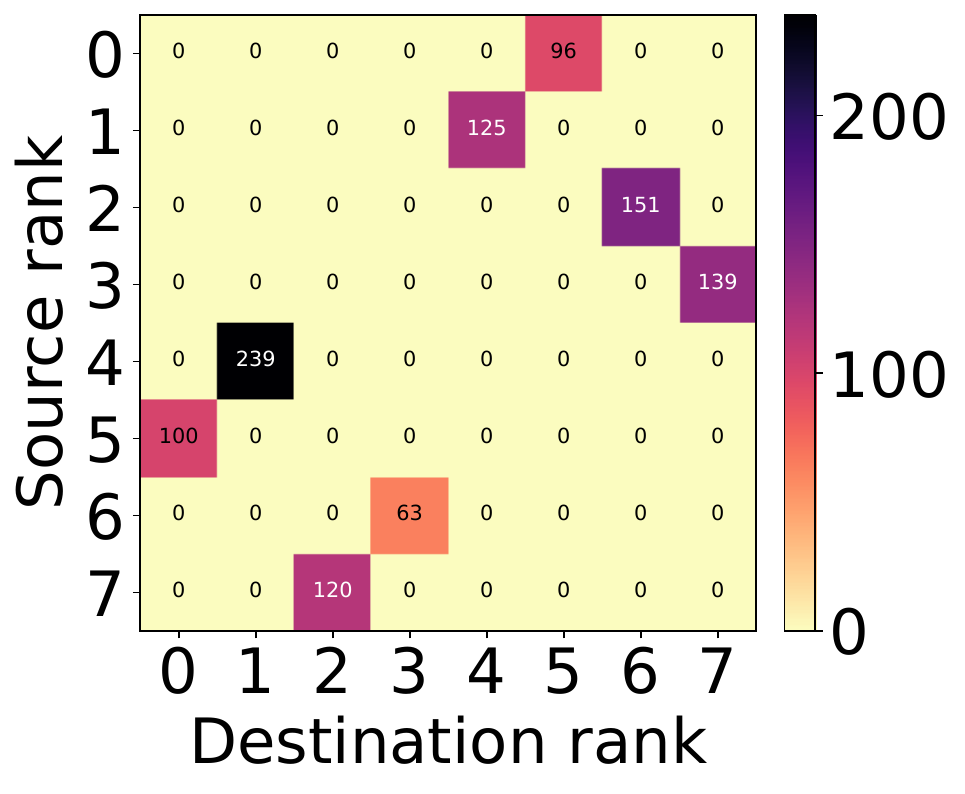}
\caption{BvN-$4/50$}
\end{subfigure}\hfill
\begin{subfigure}{0.138\linewidth}
\centering
\includegraphics[width=1\linewidth]{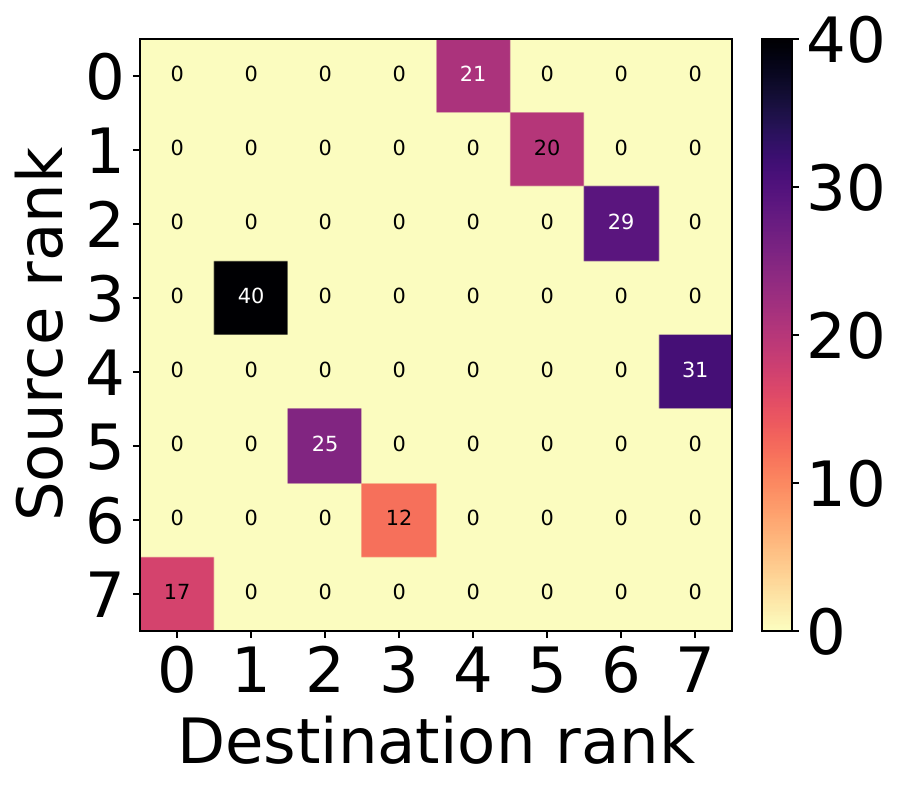}
\caption{BvN-$5/50$}
\end{subfigure}\hfill
\begin{subfigure}{0.138\linewidth}
\centering
\includegraphics[width=1\linewidth]{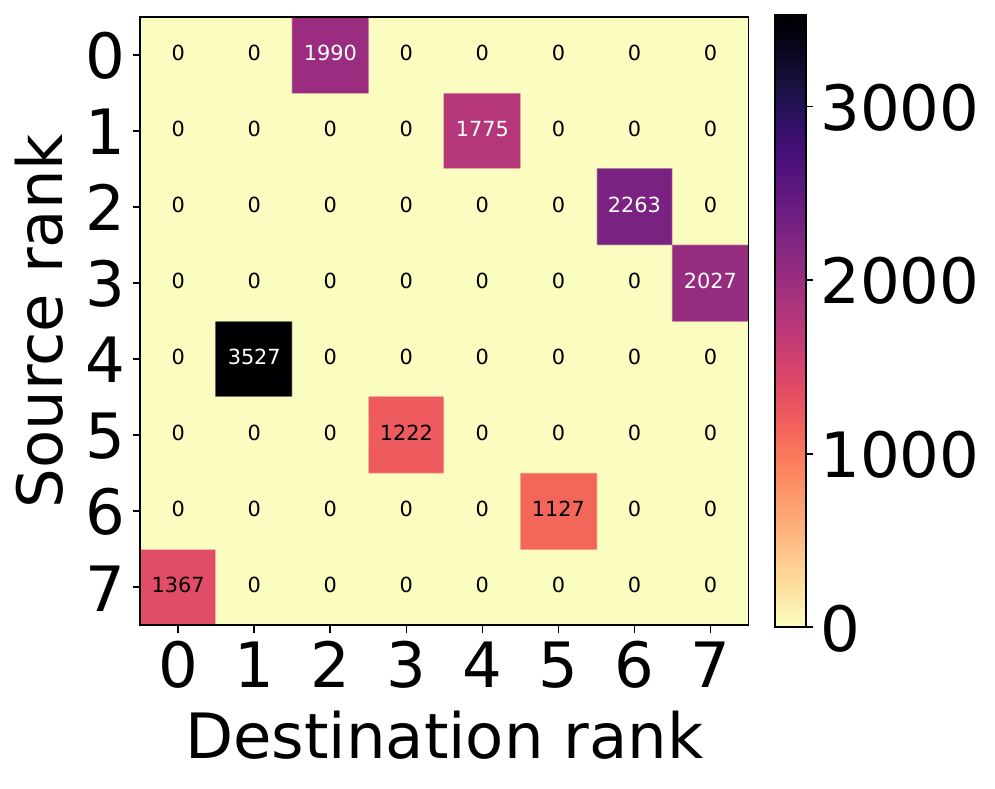}
\caption{MW-$1/7$}
\end{subfigure}\hfill
\begin{subfigure}{0.138\linewidth}
\centering
\includegraphics[width=1\linewidth]{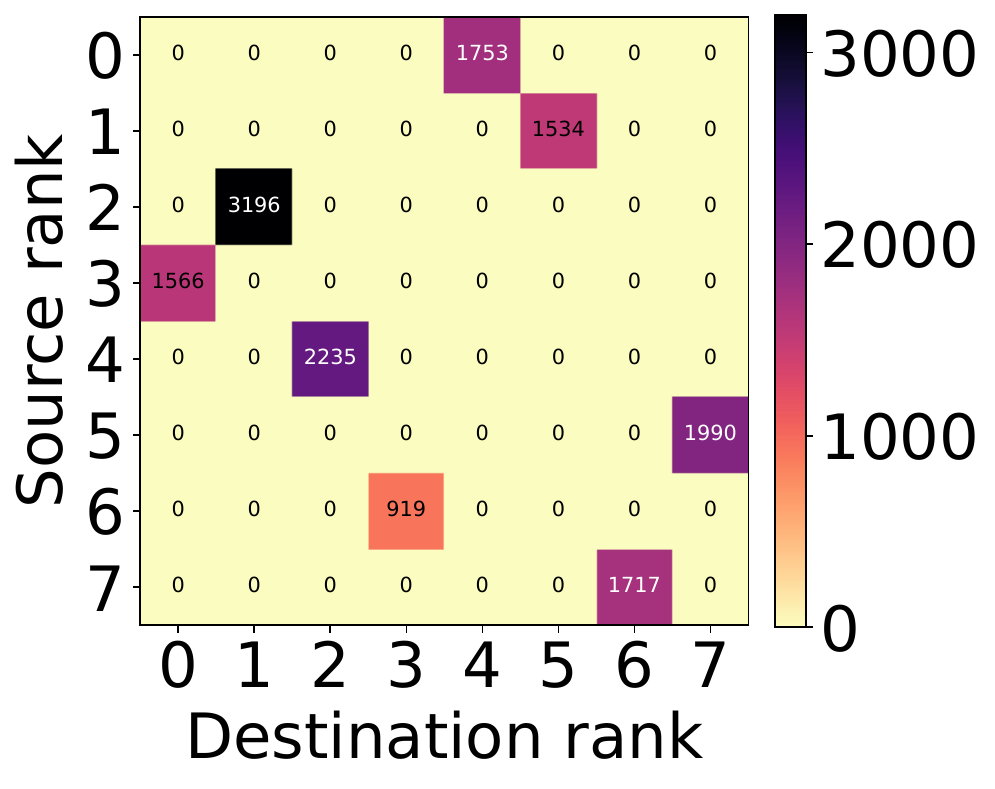}
\caption{MW-$2/7$}
\end{subfigure}\hfill
\begin{subfigure}{0.138\linewidth}
\centering
\includegraphics[width=1\linewidth]{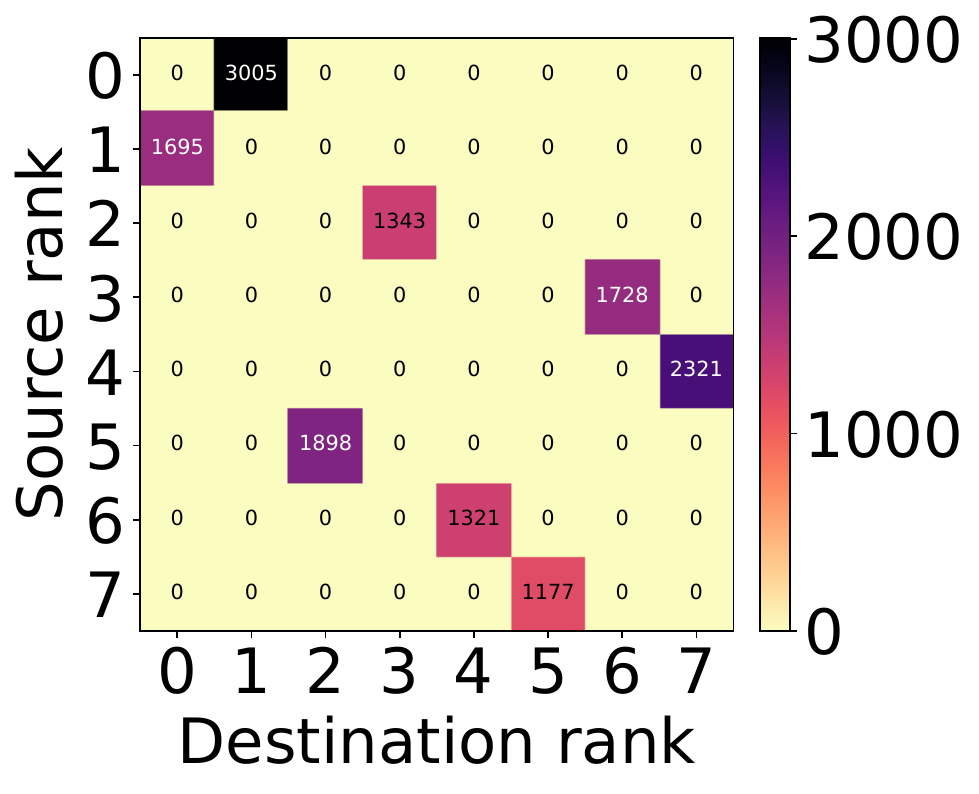}
\caption{MW-$3/7$}
\end{subfigure}\hfill
\begin{subfigure}{0.138\linewidth}
\centering
\includegraphics[width=1\linewidth]{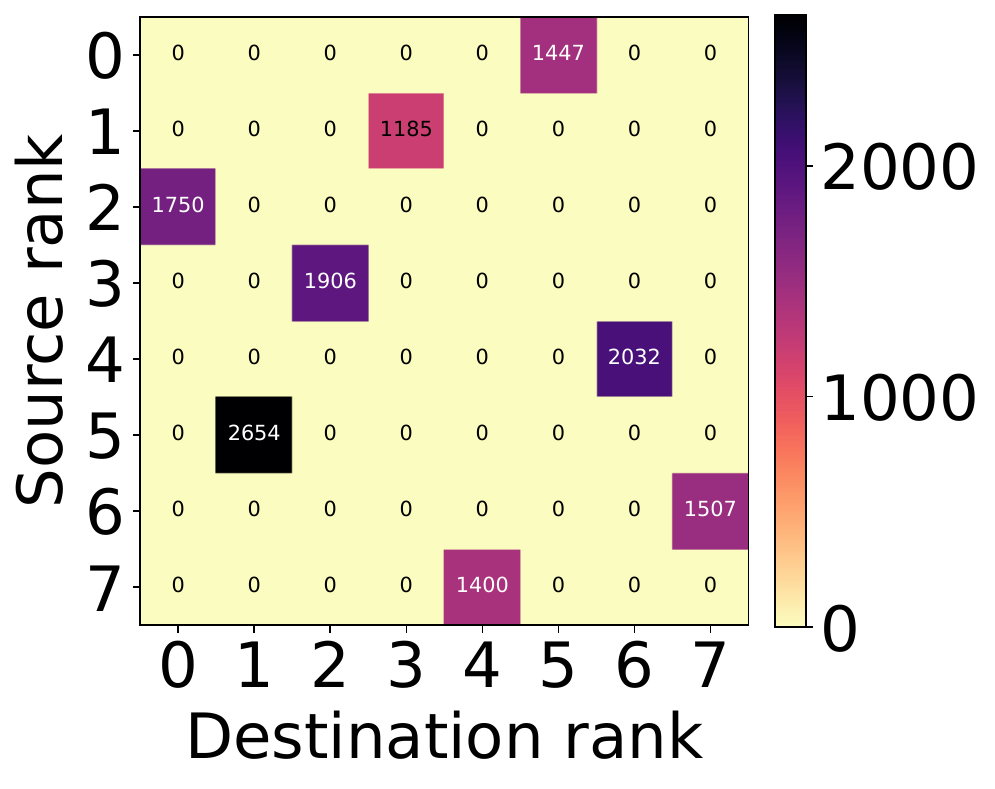}
\caption{MW-$4/7$}
\end{subfigure}\hfill
\begin{subfigure}{0.138\linewidth}
\centering
\includegraphics[width=1\linewidth]{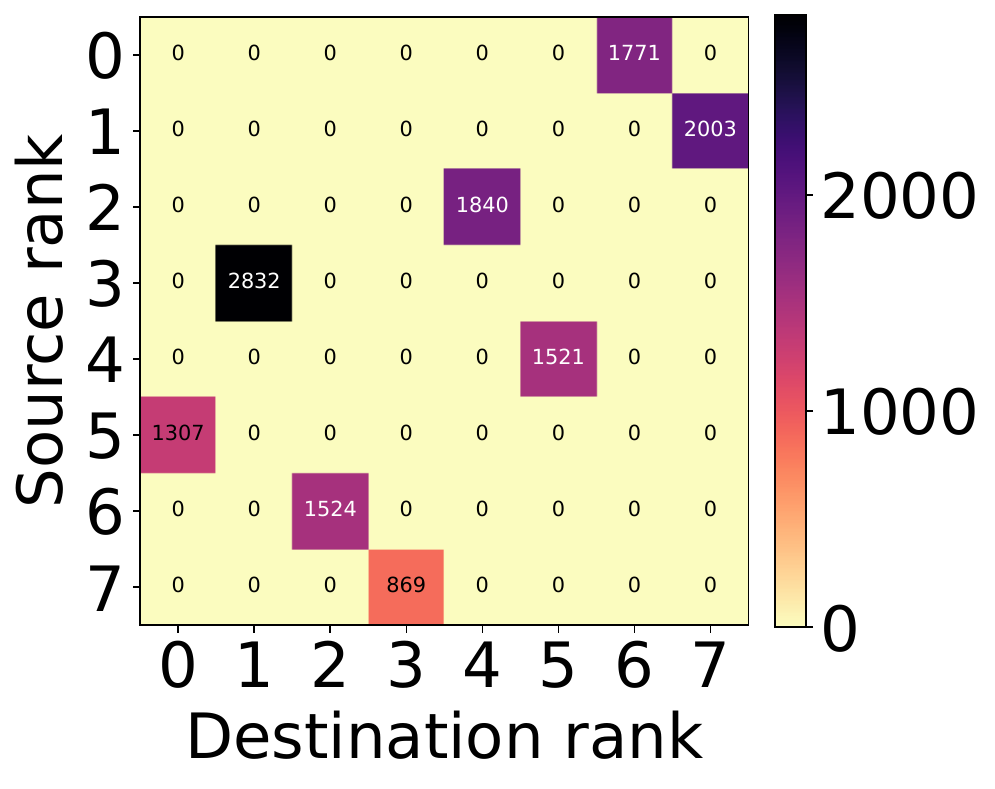}
\caption{MW-$5/7$}
\end{subfigure}\hfill
\begin{subfigure}{0.138\linewidth}
\centering
\includegraphics[width=1\linewidth]{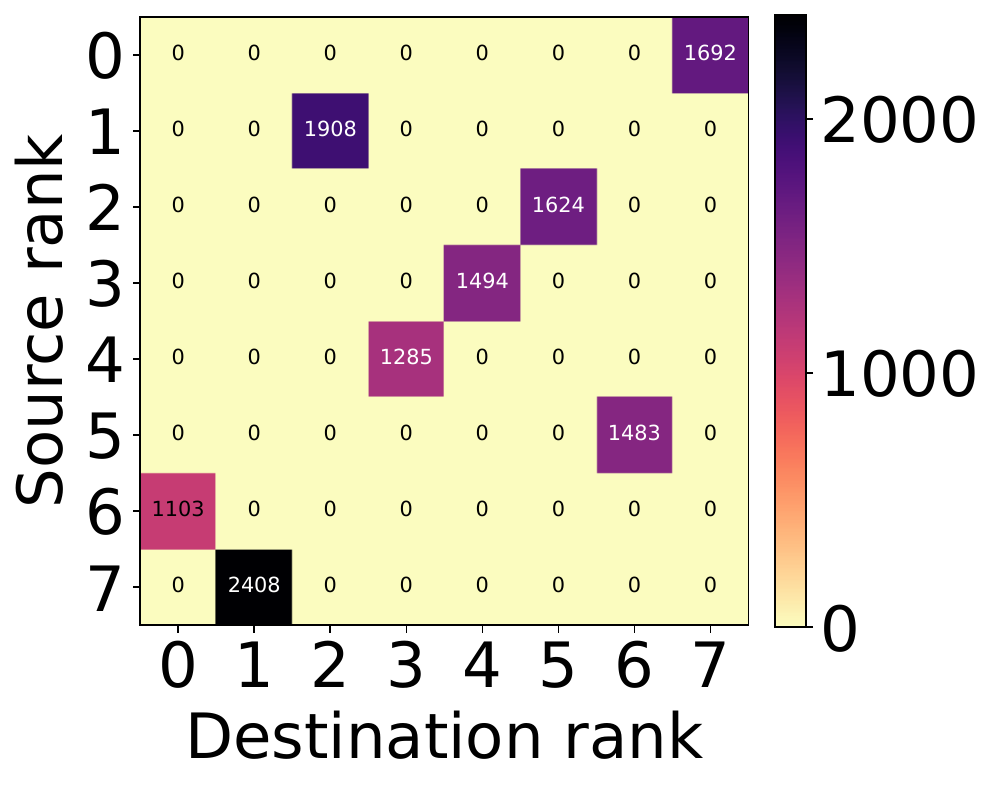}
\caption{MW-$6/7$}
\end{subfigure}\hfill
\begin{subfigure}{0.138\linewidth}
\centering
\includegraphics[width=1\linewidth]{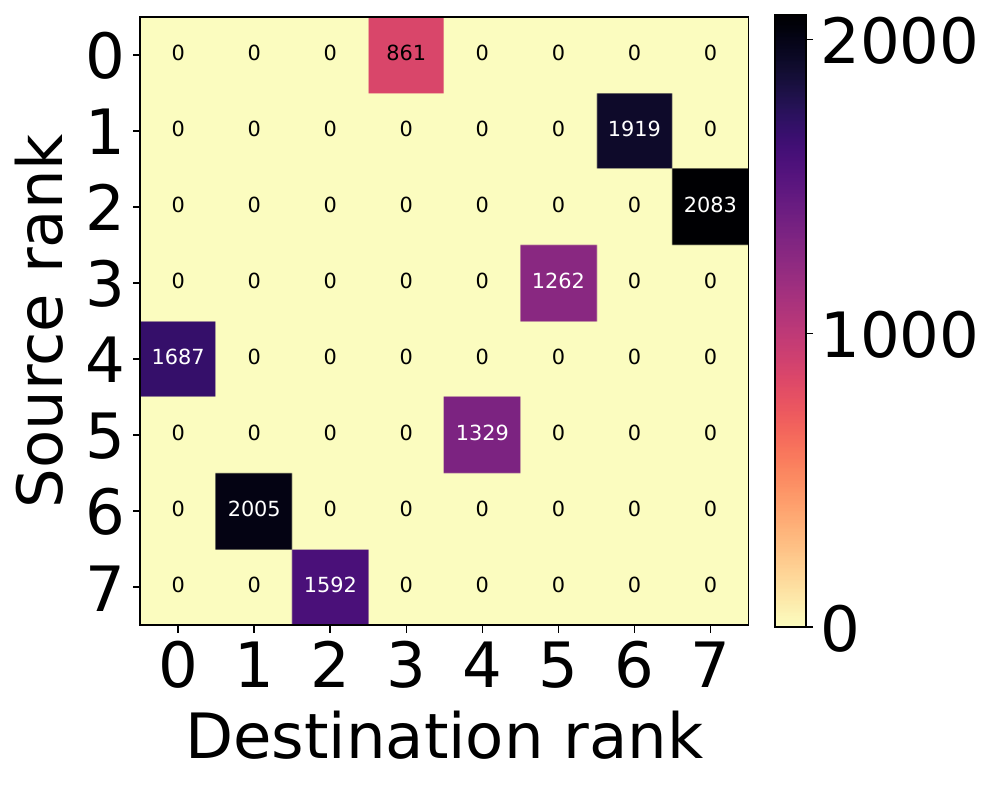}
\caption{MW-$7/7$}
\end{subfigure}\hfill
\caption{BvN decomposition results in a large number of matchings with small token counts, causing significant compute overheads for MoE execution. Whereas, the max-weight decomposition (MW) bounds the number of matchings to $O(n)$ and maintains large token counts per-matching. Note the colorbar scale.}
\label{fig:bvn-mw}
\Description{Series of heat map figures describing the sparse, and skewed routing of MoE token routing. Traces were gathered from a real MoE workload. For an $8\times 8$ matrix, BvN computed 50 matchings, while the Max Weight decomposition produces 7 matchings.}
\vspace{-4mm}
\end{figure*}

\section{Motivation}
\label{sec:motivation}
We present a brief background and motivate the need for rethinking circuit scheduling techniques in the context of distributed MoE execution.
\subsection{Mixture of Experts (MoE) Execution}
\label{subsec:MoE}

Mixture-of-Experts (MoE) architectures replace dense feed-forward layers with a pool of specialized ``experts,'' where a learned routing gate dynamically selects only a sparse subset of experts per token~\cite{Fedus2022SwitchSparsity,Shazeer2017OutrageouslyLayer}. Modern MoE models employ increasingly large expert pools; for example, Qwen3 contains $128$ experts with top-$8$ routing~\cite{Yang2025Qwen3Report}, while Mixtral 8$\times$7B employs $8$ experts with top-$2$ routing~\cite{Jiang2024MixtralExperts}. This sparsity naturally enables expert parallelism (EP), where experts are distributed across multiple GPUs or machines.

Consequently, MoE execution introduces a tightly coupled dispatch--compute--combine structure:
\begin{itemize}[leftmargin=*]
    \item[$\blacksquare$] \textbf{Dispatch:} routed tokens are exchanged across devices through all-to-all (A2A) communication to reach their assigned experts.
    \item[$\blacksquare$] \textbf{Compute:} each expert processes its assigned tokens.
    \item[$\blacksquare$] \textbf{Combine:} processed tokens are returned to their originating devices through another A2A communication.
\end{itemize}

The backward pass mirrors this structure, resulting in four all-to-all communications per MoE layer during training. Unlike dense Transformer workloads dominated by collective operations such as AllReduce, MoE execution introduces substantial all-to-all communication together with strict sequential dependencies between communication and expert computation. As a result, the efficiency of MoE execution depends not only on communication performance, but also on how effectively communication and expert computation can be overlapped.

\subsection{Communication \& Compute Interaction}

Interestingly, the communication matrices generated by MoE execution possess many of the properties that motivate reconfigurable circuit scheduling. Because expert assignments are determined dynamically by the routing gate, MoE traffic matrices are often sparse, highly skewed, and imbalanced, varying significantly across iterations and inputs. As a result, distributed MoE execution becomes a natural target for reconfigurable optical interconnects.

The decomposition techniques already employed in OCS scheduling can therefore be directly applied to MoE communication matrices, transforming the all-to-all communication into a sequence of scheduled matchings. At first glance, this appears highly attractive: decomposition not only adapts the network to the skewed communication structure of MoE workloads, but also creates opportunities to pipeline communication, expert computation, and network reconfiguration overheads\cite{Huang2016Sunflow:Coflows}.

However, decomposition does more than define the communication schedule of the network. In MoE execution, each matching also implicitly determines the granularity at which expert computation is exposed. Consider a decomposition that partitions a large all-to-all exchange into many small matchings. While such a schedule may improve instantaneous network utilization and reduce contention, each matching now carries only a small number of routed tokens per expert. Since experts may begin computation immediately upon receiving tokens, the resulting expert executions become increasingly short and fragmented.

\myitem{Compute time is not strictly linear:} Expert execution across token batch sizes typically exhibits a ``knee'' behavior (Figure~\ref{fig:compute-times}). Beyond a sufficiently large batch size, execution scales approximately linearly with the number of tokens. However, at small batch sizes, fixed kernel launch, synchronization, and scheduling overheads dominate execution time, causing compute efficiency to collapse. Kernel level optimisations remain an active area of research \cite{Aimuyo2025FlashMoE:Kernel, Chang2024FLUX:Fusion}. Nevertheless, highly fragmented decompositions may produce communication-efficient schedules that are computationally inefficient.

\myitem{Communication and compute create competing objectives:} Fine-grained decompositions improve network utilization, reduce contention, and expose additional opportunities for communication--compute overlap. At the same time, they shorten expert execution windows and reduce the amount of computation available to hide subsequent communication and reconfiguration overheads. When expert computation becomes too short, the pipeline between communication and compute breaks down, creating scheduling bubbles where both network and compute resources remain underutilized.

Consequently, decomposition quality for communication is fundamentally different from decomposition quality for MoE execution. Effective schedules must jointly balance communication efficiency, overlap opportunities, and execution granularity rather than optimizing network utilization alone.

\subsection{Scheduling Challenge in MoE Execution}

These observations fundamentally reshape the role of decomposition in distributed MoE execution. Traditionally, decomposition is viewed purely as a communication primitive whose objective is to improve network utilization and reduce contention. However, in MoE execution, decomposition simultaneously determines the ordering, granularity, and overlap structure of downstream expert computation.

This tension becomes particularly pronounced for decomposition strategies that produce a large number of fine-grained matchings. While such schedules may appear attractive from a networking perspective, they can excessively fragment expert execution into small batches that suffer poor compute efficiency and limited overlap opportunities. Furthermore, because MoE communication matrices are often sparse and highly imbalanced, decomposition strategies may introduce substantial idle capacity and scheduling bubbles even while optimizing communication structure.

As a result, minimizing communication congestion alone is insufficient and can even become counterproductive. A decomposition that is communication optimal may expose too little computation per matching to sustain efficient pipelined execution. Conversely, coarser decompositions may preserve sufficiently large expert batches to better amortize compute overheads and improve overlap effectiveness despite incurring higher instantaneous communication contention.

Motivated by these observations, we instead view distributed MoE execution as a scheduling problem that jointly considers communication decomposition and compute efficiency. Rather than optimizing communication in isolation, the objective becomes balancing communication efficiency, overlap opportunities, and execution granularity to minimize end-to-end makespan.

\section{Matrix Decompositions and Implications}

A natural approach to executing MoE communication over reconfigurable optical fabrics is to decompose the all-to-all communication matrix into a sequence of matchings that can be scheduled over time. In this section, we discuss two decomposition strategies and their implications for end-to-end MoE execution.

\subsection{Birkhoff--von Neumann Decomposition}

Birkhoff--von Neumann (BvN) decomposition is a well-established technique in optical circuit switching for deriving communication schedules from a traffic matrix. Given a doubly stochastic matrix, BvN expresses it as a convex combination of permutation matrices:
$
    A = \sum_i \lambda_i P_i,
$
where each permutation matrix $P_i$ corresponds to a perfect matching. In the context of MoE execution, applying BvN to the all-to-all communication matrix directly produces a sequence of communication matchings between ranks.

However, BvN imposes a strong structural assumption: the input matrix must be doubly stochastic. In practice, MoE communication matrices are highly sparse, skewed, and imbalanced, making them far from bistochastic. Consequently, a preprocessing step such as Sinkhorn--Knopp normalization is required before decomposition can be applied.

Figure~\ref{fig:bvn-mw} illustrates this process. Starting from the original MoE communication matrix, Sinkhorn normalization redistributes communication mass across rows and columns to satisfy bistochastic constraints. While this enables BvN decomposition, it also alters the structure of the original communication demand and introduces artificial balancing into the schedule. The resulting decomposition often produces a large number of highly fragmented matchings, many of which carry only a small number of routed tokens.

From a networking perspective, these schedules are attractive because they reduce contention and produce perfectly incast- and outcast-free communication patterns. However, this same fragmentation creates significant execution challenges in MoE workloads.

\myitem{Normalization introduces scheduling bubbles:} Sinkhorn normalization redistributes communication mass to satisfy bistochastic constraints, introducing idle capacity that does not correspond to the original routing demand. As a result, some communicating pairs finish quickly while others continue transferring substantially larger token volumes, leaving communication bubbles within the schedule.

\myitem{Fragmentation hurts compute efficiency:} BvN often produces a large number of matchings, up to $O(n^2)$ in the worst case. While decomposition enables communication--compute overlap, excessively fragmented schedules expose only small token batches per matching. Figure~\ref{fig:compute-times} shows an approximately $250\,\mu$s minimum execution overhead incurred by small batch sizes, while Figure~\ref{fig:bvn-mw} illustrates a real BvN decomposition from Mixtral-8x22B inference where several matchings carry only tens of routed tokens. Consequently, schedules that appear communication efficient can simultaneously exhibit poor end-to-end execution efficiency due to severely fragmented expert computation.

\myitem{Small matchings amplify overlap inefficiencies:} Small token batches produce short expert execution windows that are often insufficient to hide subsequent communication and reconfiguration overheads. For instance, a matching carrying only a few tens of routed tokens may complete expert computation too quickly to overlap the communication latency of the next matching, exposing both communication and reconfiguration costs.

\medskip
\noindent
\takeaway{
BvN highlights a fundamental tension in MoE execution: communication-optimal decompositions are not necessarily execution-optimal decompositions. In MoE workloads, decomposition simultaneously determines communication structure, compute granularity, and overlap opportunities, fundamentally coupling network scheduling with downstream execution efficiency.
}

\subsection{Greedy Max-Weight Decomposition}

Motivated by the limitations of highly fragmented decompositions, we explore a simple greedy max-weight decomposition strategy in which each iteration computes a maximum-weight perfect matching using the Jonker-Volgenant algorithm~\cite{7738348}. Unlike BvN, this approach operates directly on the original MoE communication matrix without requiring bistochastic preprocessing or Sinkhorn normalization.

The algorithm repeatedly extracts the maximum-weight perfect matching from the residual matrix and subtracts the selected matching until all entries become zero. For an $n \times n$ matrix, each invocation of the Jonker-Volgenant algorithm computes a maximum-weight perfect matching in $O(n^3)$ time.

Figure~\ref{fig:bvn-mw} highlights the contrast between the two approaches. While BvN decomposes the normalized matrix into a large number of fragmented matchings carrying only small token volumes, the max-weight decomposition preserves large token counts within individual matchings and produces substantially fewer schedules overall.

Preserving larger token batches directly improves expert compute granularity and exposes longer expert execution windows capable of hiding subsequent communication and reconfiguration overheads. Furthermore, because the decomposition produces fewer matchings overall, it also reduces the number of synchronization points and topology reconfiguration events exposed during execution.

As a result, despite potentially introducing higher instantaneous communication contention, the overall execution often achieves significantly lower end-to-end makespan due to improved overlap behavior and substantially better compute efficiency.

\medskip
\noindent
\takeaway{
The key advantage of max-weight decomposition is not merely fewer matchings, but preserving large execution granularity that better aligns communication scheduling with expert compute behavior.
}

\subsection{Tradeoffs in Decomposition Granularity}

While the max-weight decomposition preserves larger token batches and reduces fragmentation, it also introduces a different tradeoff: individual matchings may themselves become highly imbalanced.

In particular, a single matching may contain communication pairs carrying vastly different token volumes. Since the completion time of a matching is determined by its most heavily loaded communication pair, smaller transfers may finish significantly earlier and remain idle while waiting for the bottleneck transfer to complete. Consequently, even though the decomposition produces fewer and denser matchings overall, imbalance within a matching can still introduce communication bubbles and underutilization.

This effect differs fundamentally from the fragmentation observed in BvN decomposition. Rather than producing many small matchings, the max-weight decomposition concentrates communication into a smaller number of dense matchings dominated by a subset of large transfers, as illustrated in Figure~\ref{fig:bvn-mw}.

The decomposition additionally exposes a scheduling challenge regarding the ordering of matchings during execution. Large matchings expose long expert execution windows capable of hiding subsequent communication and reconfiguration overheads, while smaller residual matchings are more likely to leave the network exposed. Viewed from this perspective, the dispatch--compute--combine structure of MoE execution naturally resembles a three-machine flow-shop scheduling problem~\cite{Johnson1954OptimalIncluded}, where the objective is to minimize end-to-end makespan through the joint optimization of matching order and overlap behavior.

\medskip
\noindent
\takeaway{
Decomposition quality is not solely determined by the number of matchings produced. The internal structure, imbalance, and ordering of matchings fundamentally shape overlap opportunities and end-to-end execution efficiency.
}

\begin{figure*}
\centering
\begin{subfigure}{1\linewidth}
\centering
\includegraphics[width=0.4\linewidth]{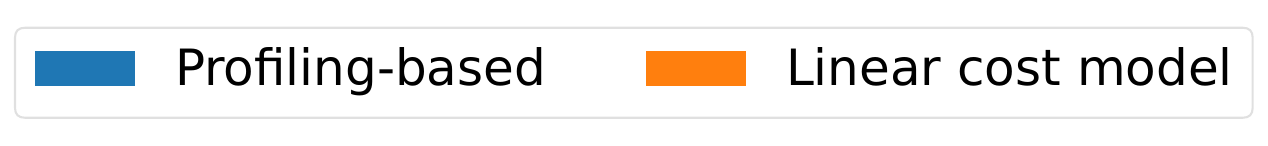}
\end{subfigure}
\begin{subfigure}{0.32\linewidth}
\centering
\includegraphics[width=1\linewidth]{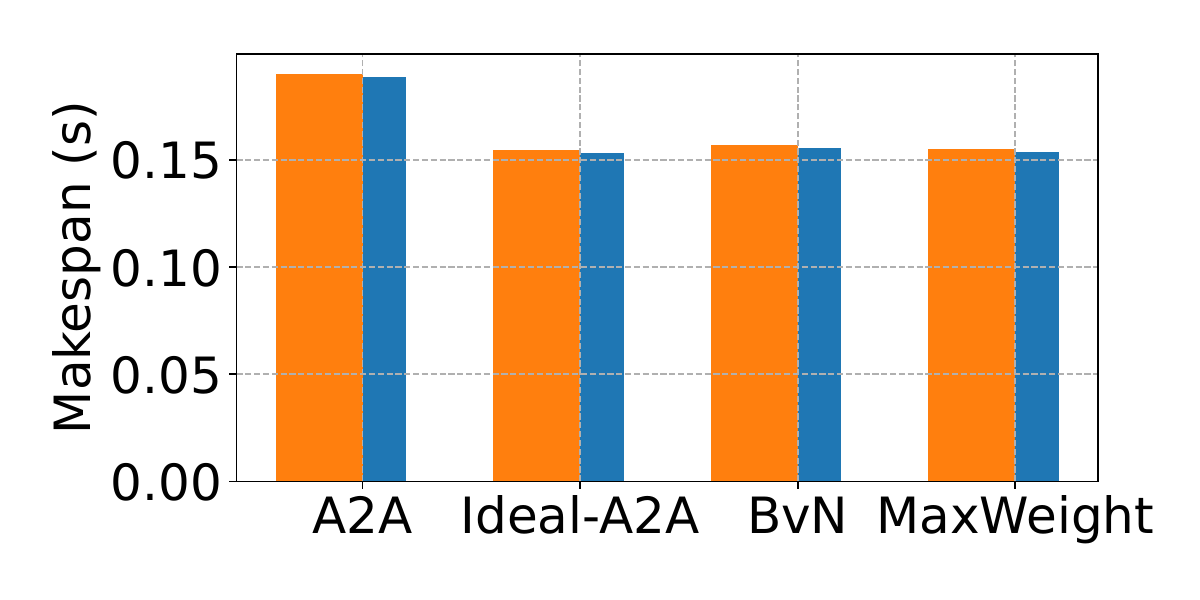}
\caption{Mixtral 8x7b, no overlap}
\end{subfigure}
\begin{subfigure}{0.32\linewidth}
\centering
\includegraphics[width=1\linewidth]{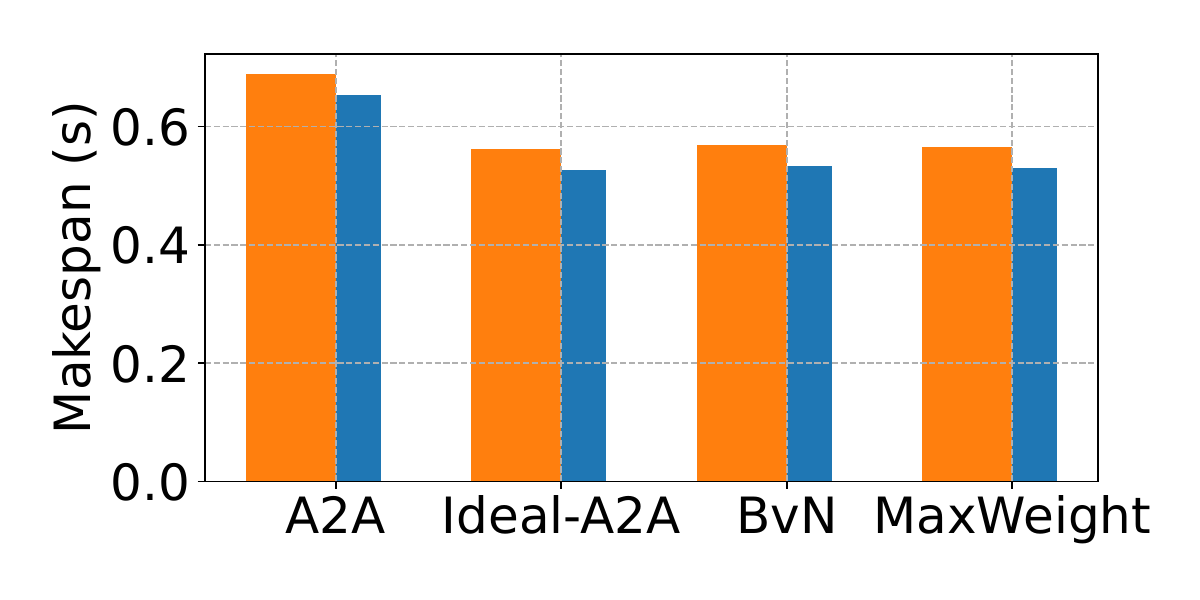}
\caption{Mixtral 8x22b, no overlap}
\end{subfigure}
\begin{subfigure}{0.32\linewidth}
\centering
\includegraphics[width=1\linewidth]{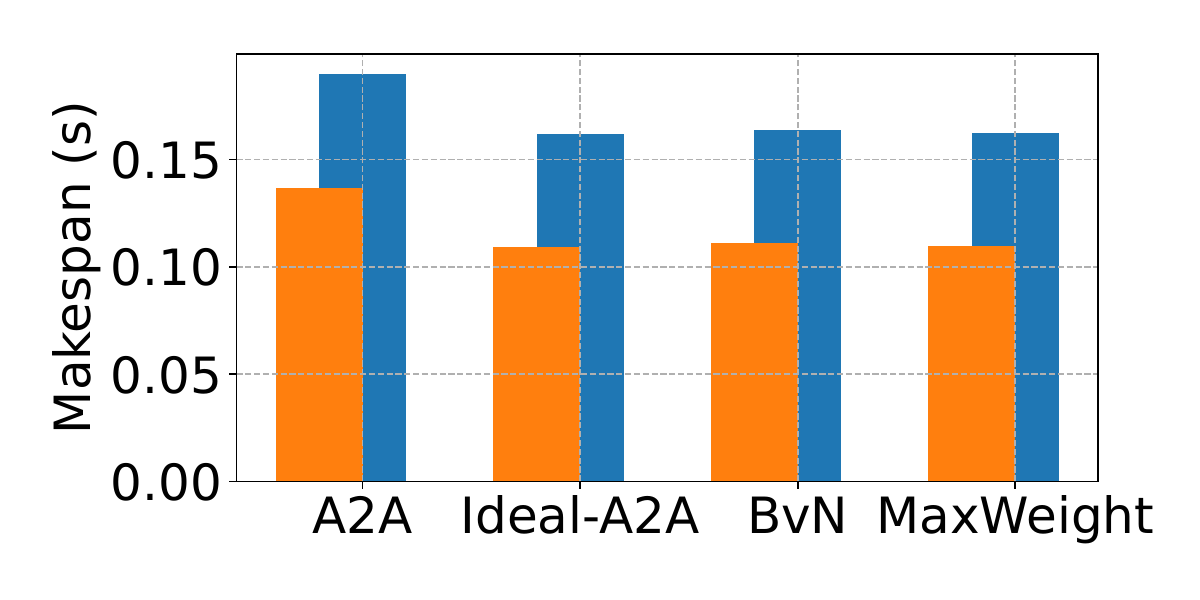}
\caption{DeepSeek MoE 16b, no overlap}
\label{fig:deepseek_moe_16b_no_overlap}
\end{subfigure}
\begin{subfigure}{0.32\linewidth}
\centering
\includegraphics[width=1\linewidth]{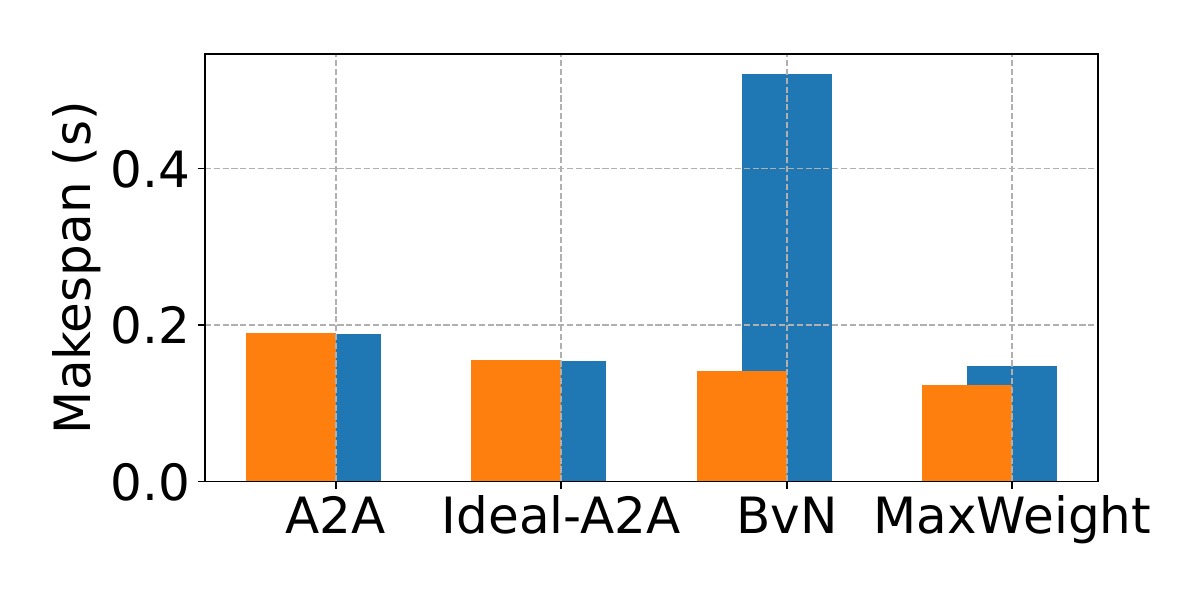}
\caption{Mixtral 8x7b, with overlap}
\end{subfigure}
\begin{subfigure}{0.32\linewidth}
\centering
\includegraphics[width=1\linewidth]{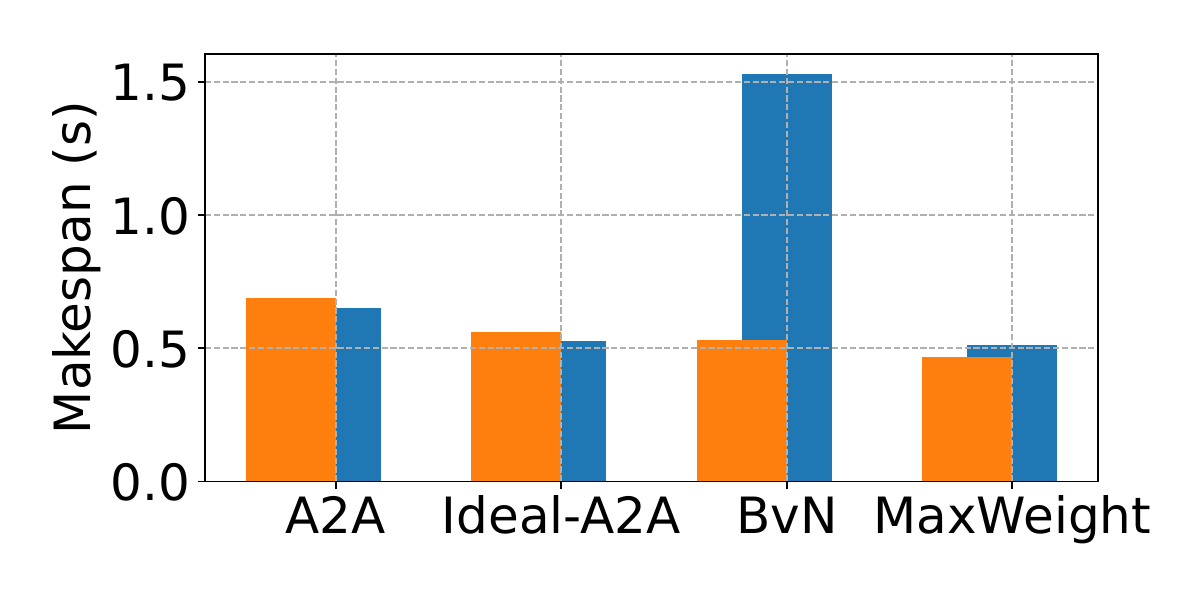}
\caption{Mixtral 8x22b, with overlap}
\end{subfigure}
\begin{subfigure}{0.32\linewidth}
\centering
\includegraphics[width=1\linewidth]{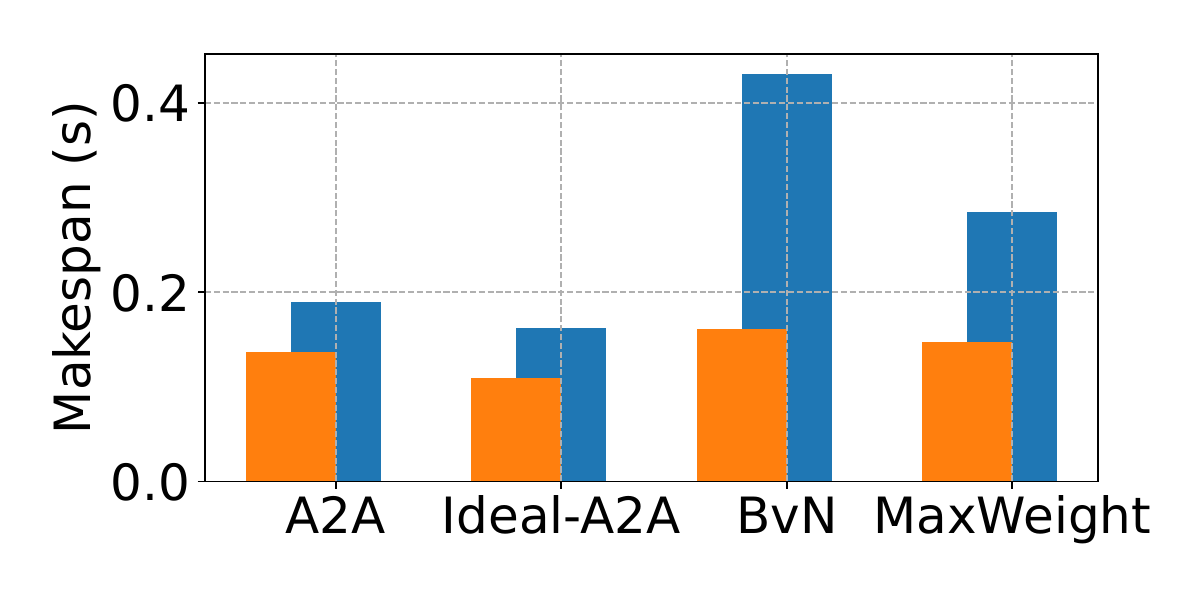}
\caption{DeepSeek MoE 16b, with overlap}
\end{subfigure}
\caption{Makespan of the MoE forward pass under different decomposition strategies on the MMLU dataset, whose small prompt sizes often lead to small effective batch sizes during execution. Neither BvN nor greedy max-weight decompositions perform well, even under near-zero reconfiguration delay, compared to standard all-to-all communication over a static ring topology despite its severe congestion.}
\label{fig:eval1}
\Description{Comparison image of the MoE makespan under different decomposition strategies on the MMLU dataset. Each image shows the performance of each decomposition strategy under the profiling-based and linear cost model. Each image contains four decomposition strategies: All-to-all, ideal All-to-all, BvN, and MaxWeight. The top set of images measures the performance where the MoE operations are not overlapped while the bottom set measure the performance of overlapped MoE operations. The decomposition based strategies do not significantly outperform the congestion prone all-to-all because of the small batches and fixed compute overheads observed in Figure \ref{fig:compute-times}}
\end{figure*}

\section{Evaluation}
\label{sec:eval}

Our evaluation focuses on end-to-end MoE layer makespan and studies how decomposition granularity, overlap opportunities, and communication structure interact under different workload characteristics. To this end, we implement a trace-driven simulator using event queues to model the dispatch--compute--combine structure of the MoE forward pass, including all-to-all communication, expert execution, and communication--compute overlap. The simulator incorporates real routing traces and configurations from Mixtral 8$\times$7B, Mixtral 8$\times$22B, and DeepSeek MoE 16B models. 

\subsection{Setup}

\myitem{Comparisons:} We compare sequential all-to-all execution, BvN decomposition, greedy max-weight decomposition, and an idealized congestion-free all-to-all baseline representing the theoretical completion time for a given communication matrix. We consider a system of $8$ GPUs connected by a circuit-switched reconfigurable interconnect. The sequential all-to-all baseline operates over a static ring topology, while the decomposition-based strategies dynamically reconfigure circuits between communicating pairs after each matching. To isolate the impact of decomposition granularity and scheduling behavior, we deliberately assume a low reconfiguration delay of $10$ns (e.g., Sirius~\cite{10.1145/3387514.3406221}). Even under such optimistic assumptions, decomposition quality significantly impacts end-to-end makespan. In practice, larger reconfiguration delays can further influence the optimal scheduling strategy and the tradeoff between communication efficiency and execution granularity~\cite{rahman2026harvestadaptivephotonicswitching,10.1145/2896377.2901479,10.1145/2716281.2836126}, opening an interesting direction for future work on decomposition-aware circuit scheduling under non-negligible reconfiguration costs.

\myitem{Execution time:} To model expert execution costs, we run experiments on RTX PRO 6000 GPUs and profile the execution time of MoE expert computation across different token batch sizes. We refer to this configuration as the \emph{profiling-based} model in our evaluation. To additionally isolate the impact of decomposition granularity independent of hardware-specific effects, we also evaluate a synthetic linear compute cost model representing idealized compute scaling behavior. For static topologies, we use Gurobi~\cite{gurobi} to solve for the optimal all-to-all completion time under link capacity constraints. For decomposition-based strategies, we analytically compute the completion time of each matching as the maximum communication time across all communicating pairs within the matching, \ie the maximum transfer size divided by the available bandwidth.
The decomposition-based strategies naturally enable communication--compute overlap. After the dispatch phase of the $i$-th matching completes, expert computation begins immediately while communication for the $(i+1)$-th matching progresses concurrently. In the presence of overlap, the event-driven simulator interleaves communication and expert execution events, allowing communication and reconfiguration overheads of subsequent matchings to be hidden behind ongoing expert computation whenever the exposed compute window is sufficiently large. In contrast, the sequential all-to-all baseline performs communication and computation strictly to completion without overlap.

\myitem{Datasets:} We evaluate these strategies across two workload regimes. The MMLU dataset consists primarily of small prompts, resulting in small effective token batches during execution. In contrast, the SPEED-BENCH throughput dataset contains large prompts ($\approx 2$k tokens per prompt), producing substantially larger expert batches and greater opportunities for overlap and compute amortization.

\begin{figure*}
\centering
\begin{subfigure}{1\linewidth}
\centering
\includegraphics[width=0.4\linewidth]{plots/legend.pdf}
\end{subfigure}
\begin{subfigure}{0.32\linewidth}
\centering
\includegraphics[width=1\linewidth]{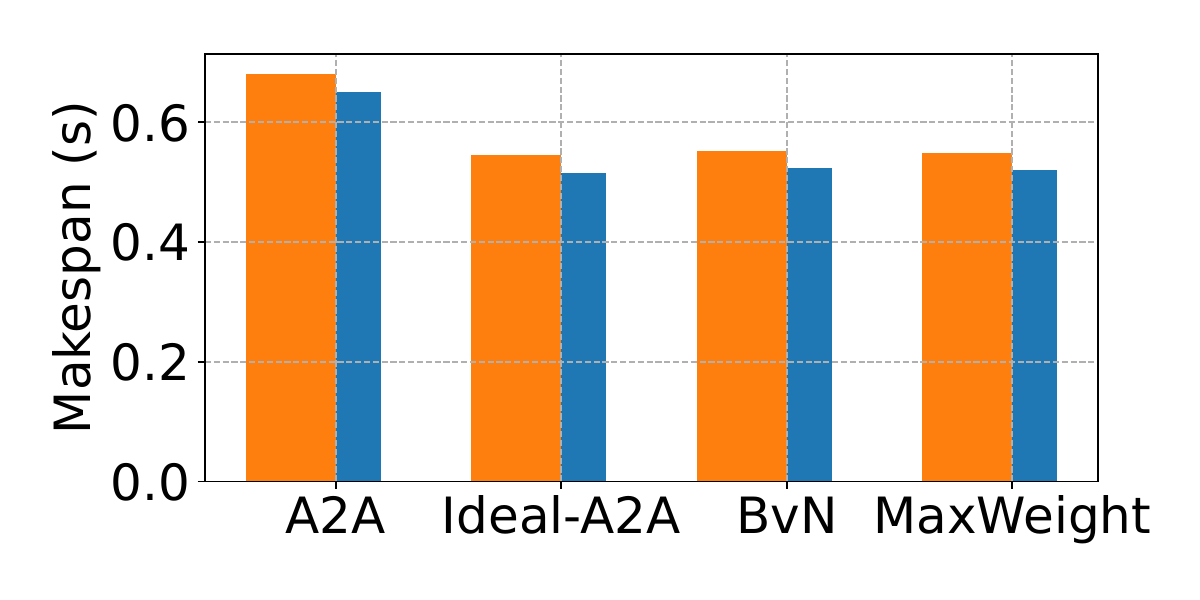}
\caption{Mixtral 8x7b, no overlap}
\end{subfigure}
\begin{subfigure}{0.32\linewidth}
\centering
\includegraphics[width=1\linewidth]{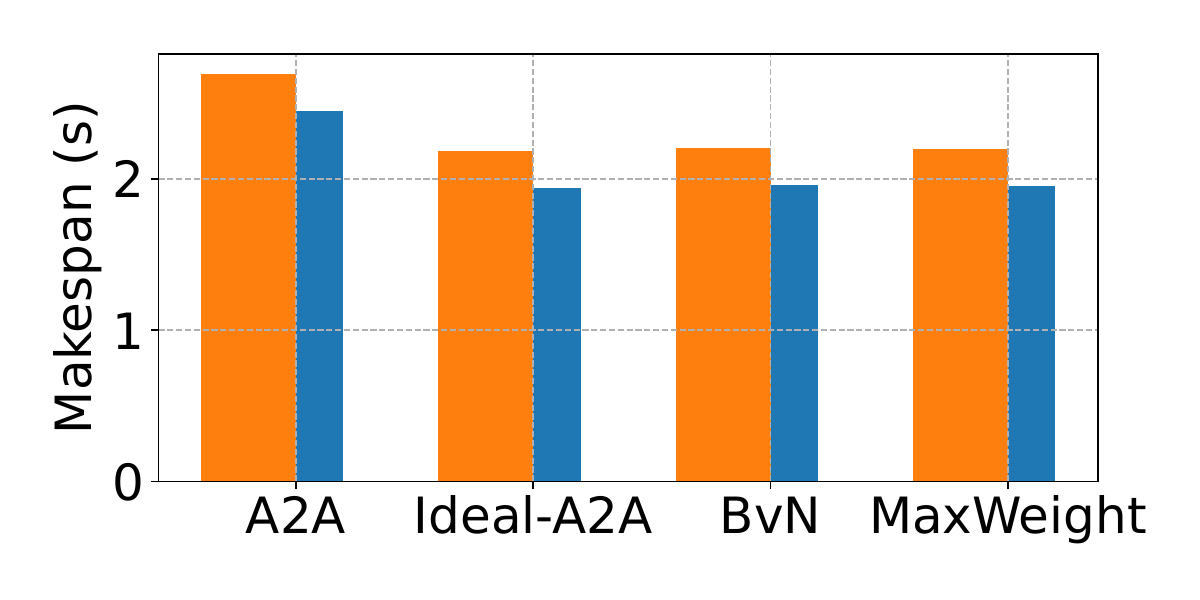}
\caption{Mixtral 8x22b, no overlap}
\end{subfigure}
\begin{subfigure}{0.32\linewidth}
\centering
\includegraphics[width=1\linewidth]{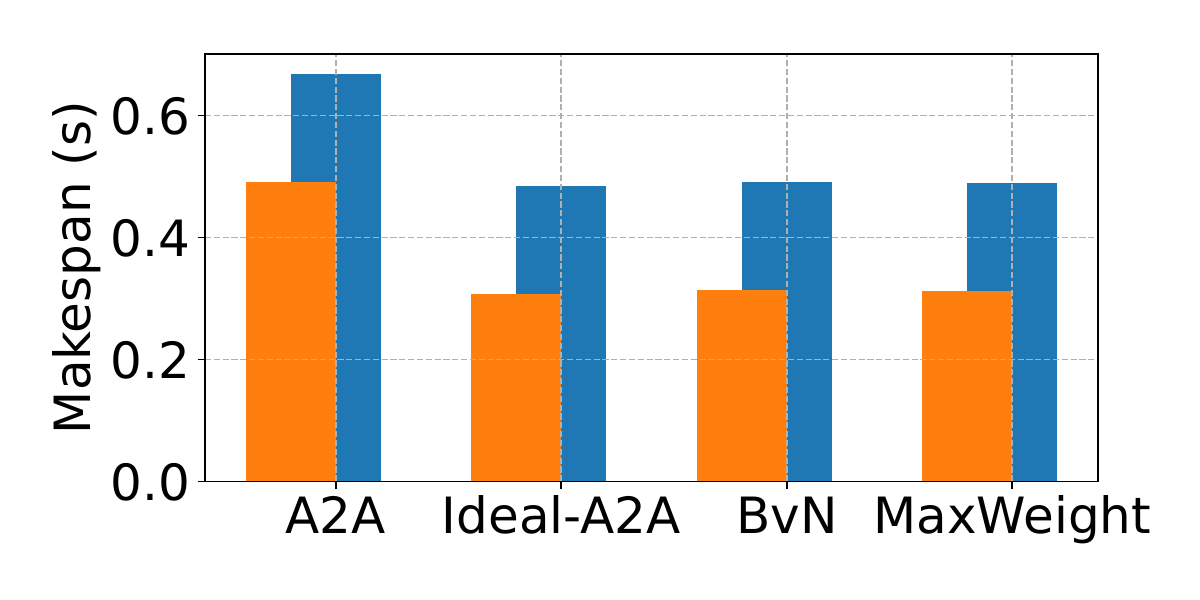}
\caption{DeepSeek MoE 16b, no overlap}
\end{subfigure}
\begin{subfigure}{0.32\linewidth}
\centering
\includegraphics[width=1\linewidth]{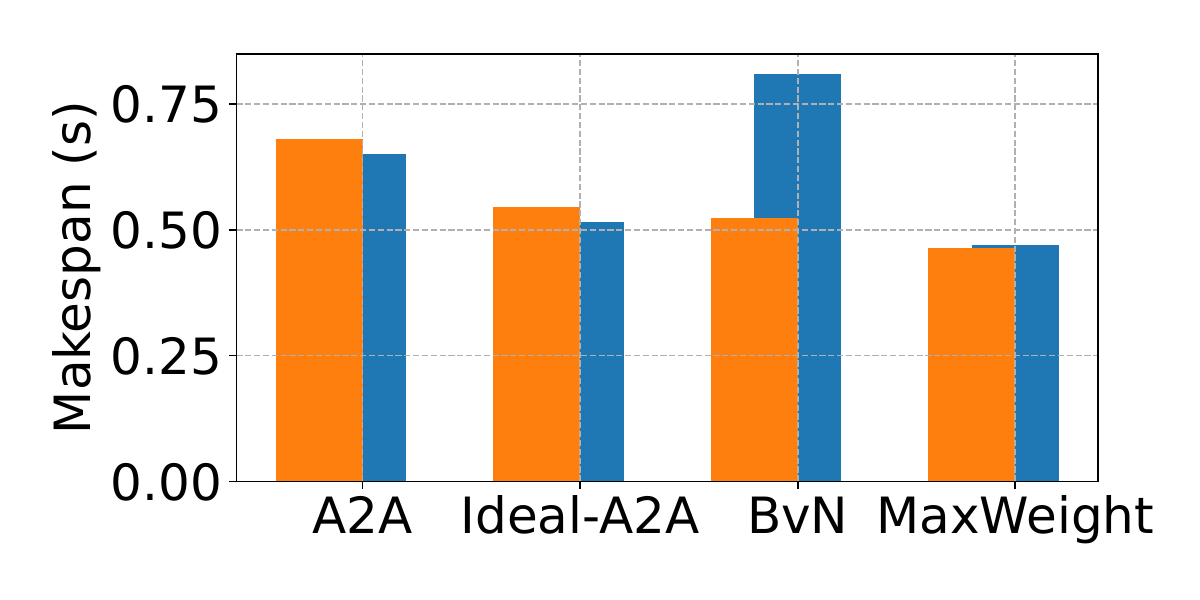}
\caption{Mixtral 8x7b, with overlap}
\end{subfigure}
\begin{subfigure}{0.32\linewidth}
\centering
\includegraphics[width=1\linewidth]{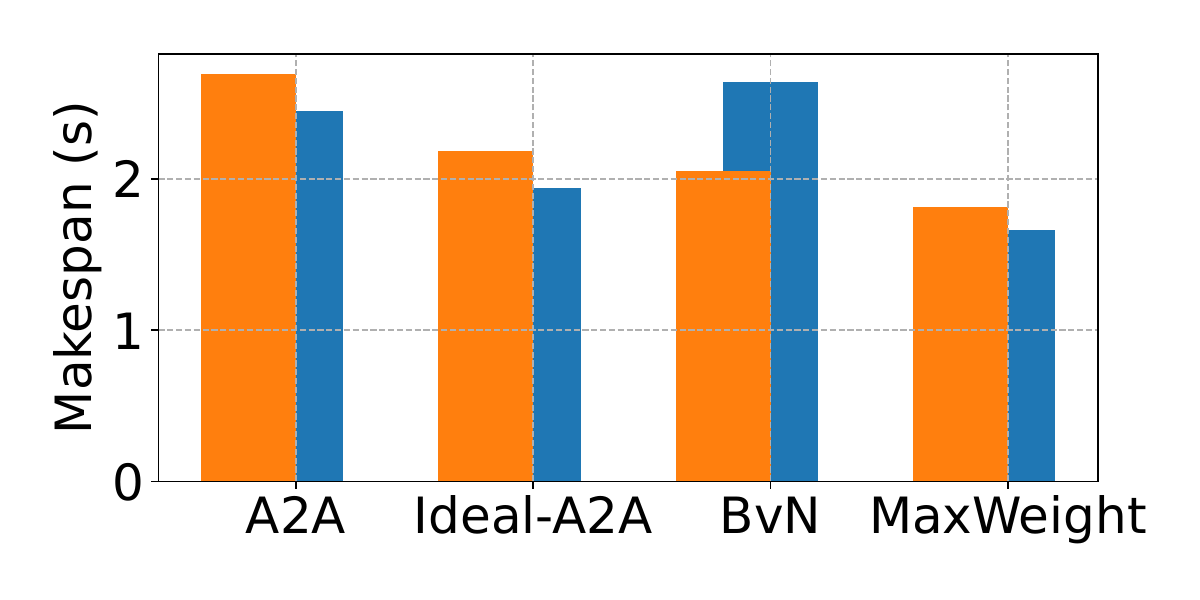}
\caption{Mixtral 8x22b, with overlap}
\end{subfigure}
\begin{subfigure}{0.32\linewidth}
\centering
\includegraphics[width=1\linewidth]{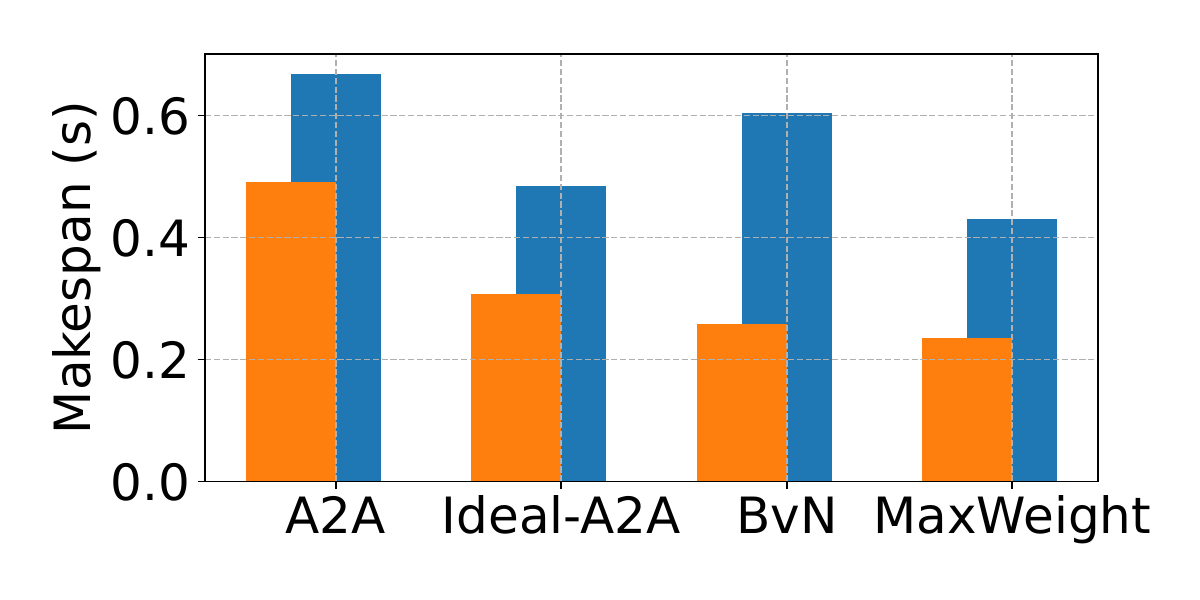}
\caption{DeepSeek MoE 16b, with overlap}
\end{subfigure}
\caption{Makespan of the MoE forward pass under different decomposition strategies on the SPEED-Bench throughput dataset, whose large prompt sizes ($\approx 2$k tokens per prompt) lead to substantially larger effective batch sizes during execution, thereby amortizing compute overheads. While BvN still suffers from the excessive number of matchings in its decomposition, greedy max-weight decomposition approaches the performance of an ideal congestion-free all-to-all and further benefits from communication--compute overlap.}
\label{fig:eval2}
\Description{Comparison image of the MoE makespan under different decomposition strategies on the SPEED-Bench dataset. Each image shows the performance of each decomposition strategy under the profiling-based and linear cost model. Each image contains four decomposition strategies: All-to-all, ideal All-to-all, BvN, and MaxWeight. The top set of images measures the performance where the MoE operations are not overlapped while the bottom set measure the performance of overlapped MoE operations. The decomposition based strategies consistently approach the ideal all-to-all benchmark. Both decomposition strategies generally outperform the ideal all-to-all benchmark with MaxWeight decomposition providing the greatest difference over BvN.}
\end{figure*}

\subsection{Results}
\myitem{BvN severely underperforms under overlapped execution:} Figure~\ref{fig:eval1} shows the results for the MMLU dataset, whose small prompt sizes lead to small effective expert batches during execution. In this regime, the fragmentation introduced by BvN decomposition becomes particularly harmful. Our profiling of Mixtral 8$\times$22B observed BvN producing up to $50$ matchings, with many coefficients around $0.03$, corresponding to only about $3\%$ of routed tokens per matching. While decomposition theoretically enables communication--compute overlap (as seen in the linear compute cost model), these extremely small batches expose insufficient expert computation to hide subsequent communication and reconfiguration overheads. As a result, overlapped BvN execution performs significantly worse than its non-overlapped counterpart due to the accumulation of exposed scheduling bubbles and compute inefficiencies.

\myitem{Static all-to-all can outperform decomposition under small batches with overlap:} Figure~\ref{fig:eval1} shows that under small-batch regimes, even a congestion-prone all-to-all over a static ring topology can outperform highly fragmented decomposition strategies. The small effective expert batches fail to amortize fixed compute overheads, causing decomposition-induced fragmentation to dominate the benefits of reduced communication contention.

\myitem{Greedy max-weight decomposition benefits significantly from overlap:} Figure~\ref{fig:eval2} shows that under large-batch regimes, decomposition-based execution becomes substantially more effective. By preserving large token counts within each matching, the greedy max-weight decomposition exposes long expert execution windows that effectively hide subsequent communication (and can potentially hide reconfiguration overheads if they are non-negligible). As a result, the strategy significantly outperforms BvN decomposition and, in several settings, even approaches or surpasses the idealized congestion-free all-to-all baseline.

\section{Conclusion}
\label{sec:conclusion}
In this paper, we showed that communication-optimal decompositions are not necessarily execution-optimal for distributed MoE workloads over reconfigurable optical fabrics. Through trace-driven evaluation, we demonstrated that preserving large execution granularity is critical for efficient communication--compute overlap, allowing simple greedy max-weight decompositions to significantly outperform fragmented schedules such as BvN. Our results suggest that decomposition should be viewed not merely as a communication primitive, but as an execution scheduling mechanism, motivating future work on decomposition-aware scheduling under non-negligible circuit reconfiguration delays.

\bibliographystyle{plainurl}
\bibliography{references}

\label{LastPage}

\label{bodyLastPage}

\end{document}